\documentclass[pss,fleqn,embeddedheads]{w-art}
\usepackage{times}
\usepackage{w-thm}
\usepackage[]{graphicx}

\begin{document}

\DOIsuffix{theDOIsuffix}
\Volume{XX}
\Issue{1}
\Copyrightissue{01}
\Month{01}
\Year{2004}
\pagespan{1}{}
\Receiveddate{\sf zzz} \Reviseddate{\sf zzz} \Accepteddate{\sf
zzz} \Dateposted{\sf zzz}
\subjclass[pacs]{73.63.Kv, 72.25.-b, 73.21.La, 72.25.Dc}

\title{Interplay of the Rashba and Dresselhaus spin-orbit coupling in 
the optical spin susceptibility of 2D electron systems}

\author[Catalina L\'opez]{Catalina L\'opez-Bastidas\footnote{Corresponding
author: email: {\sf clopez@ccmc.unam.mx}}}
\author{Jes\'us A. Maytorena}
\author{Francisco Mireles}

\address{Centro 
de Ciencias de la Materia Condensada, Universidad Nacional 
Aut\'onoma de M\'exico, Apdo. Postal 2681, 22800 Ensenada, 
Baja California, M\'exico}

\begin{abstract}
\vspace{0.1 in}
We present calculations of the frequency-dependent spin susceptibility 
tensor of a two-dimensional electron gas with competing Rashba and 
Dresselhaus spin-orbit interaction. It is shown that the interplay between 
both types of spin-orbit coupling gives rise to an anisotropic spectral 
behavior of the spin density response function 
which is significantly different from that of vanishing Rashba or Dresselhaus 
case. Strong resonances are developed in the spin susceptibility as a 
consequence of the angular anisotropy of the energy spin-splitting. 
This characteristic optical modulable response may be useful to 
experimentally probe spin accumulation and spin density currents in 
such systems.
\end{abstract}

\maketitle

\renewcommand{\leftmark}{C. L\'opez-Bastidas et al.:Interplay of the Rashba and Dresselhaus spin-orbit coupling}


Electrical manipulation of the electron and hole spins without 
the need of ferromagnetic materials and/or external magnetic 
fields is nowadays one of the central aspects in the field of 
spintronics. \cite{Wolf, Rashba_PhysE,Zutic} The presence of a sizeable 
spin-orbit interaction (SOI) in low-dimensional semiconductor 
structures and its modulation possibility (through electrical 
gating) make it a very prominent mechanism for the  access and 
manipulation of the carriers spin states.

It has been established that the dominant contributions to the SOI in 
quasi- two dimensional electron  gases (2DEG) are the 
so called Rashba and Dresselhaus SO couplings.\cite{Winkler-book}
The former results 
from the asymmetry of the confining potential that creates the 2DEG,
while the latter arises due to the inversion asymmetry of the bulk.
Several interesting effects and spin-based devices relying in these 
SOI mechanisms have been predicted and proposed in the last few years. 
For instance, the celebrated spin transistor proposed by Datta and Das 
\cite{DattaDas}, and its recent non-ballistic version \cite{Egues}. 
An intrinsic spin Hall effect in which a transverse spin current is 
driven by a dc electric field (without a net charge current)
has been also predicted to occur due to SOI effects.
\cite{Sinova,Sinova_SSC,Schliemann_IJMPB}
More recently, a spin (Hall) accumulation has been observed through optical
measurements\cite{Wunderlich,Kato,Sih}, and  lately, a purely electrical detection of 
a spin Hall current has been reported. \cite{Valenzuela}
Electric-field-induced spin orientation 
in SOI coupled systems \cite{Edelstein,Magarill,Chaplik,Kato_PRL,Nature-loss}
and strained semiconductors has been also explored.
\cite{Cheche}


On the other hand, the spin-splitting caused by SOI in electron systems 
opens the possibility of resonant effects via transitions between the 
spin-split states as a response to alternating electric fields. 
\cite{Halperin,Zhang,cond-mat,Entin,Rashba,Sipe} 
The importance of the study of such SOI effects in the dynamical regime 
(frequency dependent response) has been emphasized by
several authors studying a variety of physical aspects, including 
spin and charge optical conductivities \cite{cond-mat}, optical absorption 
spectra \cite{Sipe,Xu05}, optical control of the spin Hall current
through intense ac probing fields \cite{Wang}, electron-electron
interaction effects \cite{Finkelstein,Dimitrova}, electron-phonon 
interaction on spin Hall currents \cite{Grimaldi}, plasmon modes 
\cite{Entin,Xu03,XFWang,Pletyukhov}, and the relation between the spin Hall
conductivity and the spin susceptibility \cite{Finkelstein,Dimitrova,
Erlingsson} or the dielectric function \cite{Rashba}.

The spin susceptibility
plays a central role of the spin dynamics in a 2DEG. It 
gives the average spin polarization induced via
electric-dipole or magnetic-dipole interactions. Thus, it can be used to
obtain a magnetic susceptibility \cite{Finkelstein} or the electric-field-induced
spin orientation factor.\cite{Chaplik,Rashba_JS} Moreover, other transport
properties like charge or spin Hall conductivities can also be expressed
in terms of such spin response function.\cite{Erlingsson} 

Following S. I. Erlingsson et al.\cite{Erlingsson}, in this paper   
we report on the analytical and numerical calculations of the 
frequency-dependent spin susceptibility tensor
of a 2DEG with Rashba and Dresselhaus SOI. 
In Ref.\,\cite{Erlingsson} expressions for the tensor components
were obtained, however only approximated results were
reported in the finite frequency regime. Their analytical expressions for 
the spin susceptibilities are valid  as long as $k_{SO}/k_F <<1$ and 
$\alpha<<\beta$, where $k_{SO}$ and $k_F $ are the characteristic 
spin-orbit coupling and Fermi wave numbers, while  $\alpha$ and $\beta$ 
are the Rashba and Dresselhaus SOI strength parameters, respectively.

Here we show that in the more general case, particularly when there is a 
strong interplay between the Rashba and Dresselhaus SOI, very distinctive 
features of the optical spin susceptibility spectra arises in the system.
This suggests that an optically modulable spin density response
may be achievable in such systems. Furthermore, the calculated spectra show
that the combination of the excitation at finite frequency and the interplay 
between the Rashba and Dresselhaus couplings could also be used for measuring 
the ratio between the SO coupling parameters.



We consider a 2D free electron system lying at the $z=0$ plane subjected to
spin-orbit interaction, with a Hamiltonian given by 
\begin{equation} \label{H}
H=\frac{\hbar^2(k_x^2+k_y^2)}{2m^*}\,\mbox{{\bf I}}+
\alpha(k_x\sigma_y-k_y\sigma_x)+\beta(k_x\sigma_x-k_y\sigma_y)\ \ , 
\end{equation}
where $k_x, k_y$ are the components of the 2D electron wave vector, 
{\bf I} is the $2\times 2$ unit matrix and $\sigma_{\mu}$ are the Pauli matrices. 
The second term corresponds to the Rashba SO coupling which originates 
from any source of structural inversion asymmetry (SIA) of the confining potential. 
The third term is the linear Dresselhaus coupling which results from 
bulk-induced inversion asymmetry (BIA) in a narrow [001] quantum well.
Spin-orbit interaction appears as a relativistic correction (derived from
the Dirac equation) to the Hamiltonian of a slow electron. It includes
the gradient of a potential in which the electron moves. In atomic physics
such term leads to the well known ${\bf L}\cdot {\bf S}$ coupling between
the orbital and intrinsic angular momentum due to the Coulomb potential. 
For an electron in a crystal environment there are several sources of
potential gradient (impurities, confinement, boundaries, external electric field) 
which lead to an enhancement of SO coupling in solids. For quasi-2D systems
the more significant contributions are those due to SIA (Rashba) and BIA 
(Dresselhaus).\cite{Winkler-book}

The eigenstates 
$|{\bf k}\lambda\rangle$ for the in-plane motion are specified by the 
wave vector ${\bf k}=(k_x,k_y)=k(\cos\theta,\sin\theta)$ and  chirality 
$\lambda=\pm $ of the spin branches.  The double sign corresponds to 
the upper (+) and lower ($-$) parts of the energy spectrum given by
\begin{equation} 
\varepsilon_{\lambda}(k,\theta)=\frac{\hbar^2}{2m^*}(k+\lambda k_{so}(\theta))^2-
\frac{\hbar^2k_{so}^2(\theta)}{2m^*}
\end{equation}
where $k_{so}(\theta)=m^*\Delta(\theta)/\hbar^2$ is the characteristic 
SO momentum, $\Delta^2(\theta)=\alpha^2+\beta^2-2\alpha\beta \sin2\theta$ 
describes the angular anisotropy of the spin splitting. At zero temperature,
the two spin-split subbands are filled up to the same (positive) Fermi energy 
level $\varepsilon_F$ but with different Fermi wave vectors  
$q_{\lambda}(\theta)=\sqrt{2m^*\varepsilon_F/\hbar^2+k_{so}^2(\theta)} 
-\lambda k_{so}(\theta)$, determined from the equations  
$\varepsilon_{\lambda}(q_{\lambda}(\theta),\theta)=\varepsilon_F$.
Here, $\varepsilon_F=\hbar^2(k_0^2-2q_{so}^2)/2m^* $ with 
$k_0=\sqrt{2\pi n}$ being the Fermi wave vector of a spin-degenerate 
2DEG with electron density $n$, and $q_{so}=m^*\sqrt{\alpha^2+\beta^2}/\hbar^2$. 
The SOI splits the Fermi line into two curves with radii 
given by $q_{\lambda}(\theta)$ which, as the energy surfaces 
$\varepsilon_{\lambda}({\bf k})$, are symmetric with respect to the 
(1,1) and (-1,1) directions in ${\bf k}-$space (Fig.\,1). 
When $\alpha$ or $\beta$ is null, the dispersions are isotropic and 
the Fermi contours are concentric circles. If $\alpha=\pm\beta$ the
spin-splitting along the ($\pm$1,1) direction vanishes and the 
spin-split dispersion branches are two circles with the same radius
and displaced from the origin (along ($\mp$1,1) direction).

\begin{figure}
\includegraphics[width=0.5\textwidth]{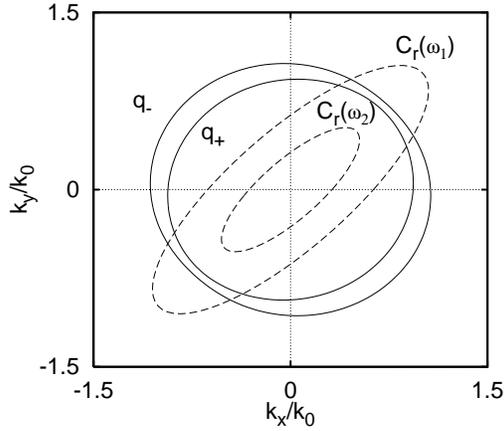}
\caption{Fermi contours $q_{\lambda}(\theta)$ and the 
constant-energy-difference curve $C_r(\omega)$ defined by 
$\varepsilon_+({\bf k})-\varepsilon_-({\bf k})=\hbar\omega$, shown 
for two values of the photon energy, $\hbar\omega_1>\hbar\omega_2$. 
$C_r(\omega)$ is a rotated ellipse 
with semi-axis of lengths (major) $k_a(\omega)=\hbar\omega/2|\alpha-\beta|$ and 
(minor) $k_b(\omega)=\hbar\omega/2|\alpha+\beta|$ oriented along the $(1,1)$ and 
$(-1,1)$ directions respectively.
The sample parameters used here are $n=5\times 10^{11}$cm$^{-2}$, 
$\alpha=1.6\times 10^{-9}\,$eV\,cm, $\beta=0.5\alpha$ and $m^*=0.055m$.}
\end{figure}

Within the linear response Kubo formalism the
spin susceptibility is given by 
\begin{equation} \label{Kubo_chi}
\chi_{\mu\mu'}(\omega)=\frac{i}{\hbar}\,
\int_0^{\infty}\!dt\,e^{i(\omega+i\eta)t}\langle[\sigma_{\mu}(t),\sigma_{\mu'}(0)]\rangle \ \ ,
\ \ \ \ \ \mu,\mu'=x,y
\end{equation}
where the symbol $\langle\cdots\rangle=\Sigma_{\lambda}\int d^2 k\, 
f(\epsilon_{\lambda}({\text \bf k}))(\cdots)$ indicates quantum and thermal 
averaging, $f(\epsilon)$ is the Fermi distribution function, and $\eta\to 0^+$. 
This is a spin-spin response function
for a spatially homogeneous (in-plane) perturbation oscillating at frequency $\omega$.

In the limit of vanishing temperature, eq. (\ref{Kubo_chi}) takes the form
\begin{equation}
\chi_{\mu\mu'}(\omega)=\frac{1}{\pi^2}\int^{\prime}\!d^2k\,
\langle -|\sigma_{\mu}|+\rangle\langle +|\sigma_{\mu'}|-\rangle\,
\frac{\varepsilon_+({\bf k})-\varepsilon_-({\bf k})}
{[\varepsilon_+({\bf k})-\varepsilon_-({\bf k})]^2-[\hbar(\omega+i\eta)]^2} \ \ ,
\end{equation}
the prime on the integral indicates that integration is restricted to 
the region between the Fermi contours, $q_+(\theta)<k<q_-(\theta)$,
for which $\varepsilon_-({\bf k})<\varepsilon_F<\varepsilon_+({\bf k})$,
(Fig. 1).

Using the result
\begin{equation}
\langle -|\sigma_{\mu}|+\rangle=-\langle +|\sigma_{\mu}|-\rangle=
\frac{i}{\Delta(\theta)}[\delta_{\mu x}(\alpha\cos\theta-\beta\sin\theta)+
\delta_{\mu y}(\alpha\sin\theta-\beta\cos\theta)]
\end{equation}
the susceptibility tensor becomes
\begin{equation} \label{chi_ij}
\chi_{\mu\mu'}(\omega)=\frac{1}{\pi^2}\,
\int_0^{2\pi}\!\!d\theta\,\frac{g_{\mu\mu'}(\theta)}{\Delta(\theta)}\,
\int_{q_+(\theta)}^{q_-(\theta)}\!\!dk\,\frac{k^2}
{4k^2\Delta^2(\theta)-[\hbar(\omega+i\eta)]^2} \ \ ,
\end{equation}
where
\begin{eqnarray}
g_{\mu\mu'}(\theta)&=&\delta_{\mu\mu'}[\delta_{\mu x}(\alpha\cos\theta-\beta\sin\theta)^2+
\delta_{\mu y}(\alpha\sin\theta-\beta\cos\theta)^2]  \nonumber \\
&&\hspace*{4cm}  +(1-\delta_{\mu\mu'})(\alpha\cos\theta-\beta\sin\theta)
(\alpha\sin\theta-\beta\cos\theta) \nonumber \ \ .
\end{eqnarray}
It can be shown that $\chi_{xx}(\omega)=\chi_{yy}(\omega)$ and 
$\chi_{xy}(\omega)=\chi_{yx}(\omega)$.
Note also that for $\beta=0$, $\,\chi_{xy}(\omega)=\chi_{yx}(\omega)=0\,$.

We can write the susceptibility in the form
$\chi_{\mu\mu'}=\chi'_{\mu\mu'}+i\chi''_{\mu\mu'}$.
The real part is
\begin{equation} \label{Rechi}
\chi'_{\mu\mu'}(\omega)=\chi_{\mu\mu'}(0)+\frac{\hbar\omega}{16\pi^2}\,
\int_0^{2\pi}\!\!d\theta\,\frac{g_{\mu\mu'}(\theta)}{\Delta^4(\theta)}\,
\log\!\left|\frac{[\omega+\Omega_+(\theta)]
[\omega-\Omega_-(\theta)]}
{[\omega+\Omega_-(\theta)][\omega-\Omega_+(\theta)]}\right|
\end{equation}
where
$\hbar\Omega_{\pm}=|\varepsilon_F-\varepsilon_{\mp}(q_{\pm}(\theta),\theta)|=
2q_{\pm}(\theta)\Delta(\theta)$ and the static value is
\begin{equation}
\chi_{\mu\mu'}(0)=\frac{\nu_0}{2}
\left[\delta_{\mu\mu'}-(1-\delta_{\mu\mu'})
\left(\frac{\beta}{\alpha}\Theta(\alpha^2-\beta^2)+
\frac{\alpha}{\beta}\Theta(\beta^2-\alpha^2)\right)\right] \ \ ,
\end{equation}
$\nu_0=m^*/\pi\hbar^2$ is the density of states of a spin-degenerate
2DEG, and $\Theta(x)$ is the unit step function, $\Theta(x)=1$ if $x>0$ and
$\Theta(x)=0$ if $x<0\,$.

For the imaginary part we have
\begin{eqnarray}
\chi''_{\mu\mu'}(\omega)&=&\pi\int'\!\!\frac{d^2k}{(2\pi)^2}\,
\langle -|\sigma_{\mu}|+\rangle\langle+|\sigma_{\mu'}|-\rangle
\,\delta(\varepsilon_+({\bf k})-\varepsilon_-({\bf k})-\hbar\omega) \label{Im_chi}\\
&=&\frac{\hbar\omega}{16\pi}
\int\!\!d\theta\,\frac{g_{\mu\mu'}(\theta)}{\Delta^4(\theta)}\,
\Theta[\hbar\omega-\hbar\Omega_+(\theta)] \,
\Theta[\hbar\Omega_-(\theta)-\hbar\omega] \label{Imchi}\ \ .
\end{eqnarray}
These equations express the fact that the only transitions allowed
between spin-split subbands $\varepsilon_{\lambda}$ due to photon
absorption at energy $\hbar\omega$ are those for which
$\hbar\Omega_+(\theta)\leq\hbar\omega\leq\hbar\Omega_-(\theta)$.
That is, for a given $\omega$ only those angles satisfying this condition
must be considered in the integral (\ref{Imchi}), see Fig.\,2c.
This is different to the pure Rashba (or Dresselhaus)
case, where the whole interval $[0,2\pi]$ contributes to the integral for each
allowed photon energy. The non-isotropic spin-splitting originated by the
simultaneous presence of both coupling strengths, forces the optical
excitation to be ${\bf k}-$selective.

\begin{figure}
\includegraphics[width=0.5\textwidth]{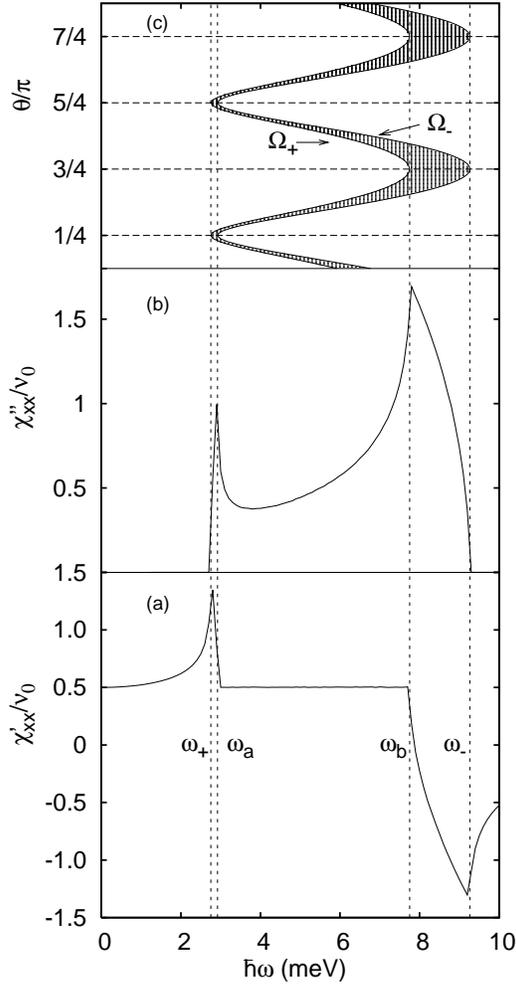}
\caption{(c) Angular region (shaded) in ${\bf k}-$space available
for direct transitions as a function of photon energy. Only the
shaded region contribute to the optical absorption [eq.\,(\ref{Imchi})]. 
The energy boundaries are given by $\hbar\Omega_+(\theta)=\varepsilon_F-
\varepsilon_-(q_+(\theta),\theta)$ and  $\hbar\Omega_-(\theta)=
\varepsilon_+(q_-(\theta),\theta)-\varepsilon_F$.
(b) Imaginary and (a) Real part of the optical spin susceptibility 
$\chi_{xx}(\omega)$, $\,\nu_0=m^*/\pi\hbar^2$. For the frequencies
$\,\omega_+=\Omega_+(\pi/4)$, $\omega_a=\Omega_-(\pi/4)$,
$\omega_b=\Omega_+(3\pi/4)$, $\omega_-=\Omega_-(3\pi/4)$, see the
text. The sample parameters are the same as in Fig.\,1\,.}
\end{figure}

In Fig.\,2 we show $\chi_{xx}(\omega)$
as obtained from eqs. (\ref{Rechi})-(\ref{Imchi}), the $xy$ component
behaves similarly. The result is
remarkably different from that of the pure Rashba or Dresselhaus case,
where the spin-splitting is isotropic in the momentum space.
For example, if $\beta=0$, $\alpha\neq 0$, then 
$\chi''_{\mu\mu'}(\omega)=\chi_R\delta_{\mu\mu'}$ only for 
$2\alpha k_+\leq\hbar\omega\leq 2\alpha k_-$, otherwise it vanishes,
where $\chi_R=\hbar\omega/16\alpha^2$, with 
$k_{\pm}=q_{\pm}(\beta=0)$ being independent of angle $\theta$;
(see Fig.\,3).
Thus, in this case the width $\Delta{\cal E}$ of the spectrum
is $\Delta{\cal E}_R=4\varepsilon_R$ (or $4\varepsilon_D$ if
$\alpha=0$, $\beta\neq 0$);
$\varepsilon_R=m^*\alpha^2/\hbar^2$ and $\varepsilon_D=m^*\beta^2/\hbar^2$
are the SO characteristic energy scales for the Rashba and Dresselhaus coupling.
As was discussed in Ref.\,\cite{cond-mat},
$\Delta{\cal E}_{R,D}$ can be about an order of magnitude
smaller than the width of the spectrum shown in Fig.\,2b.
Assuming that $\alpha>\beta$ and $(k_{so}(\theta)/k_0)^2\ll 1$, we have 
$\Delta{\cal E}=4\beta k_0+\Delta{\cal E}_R+\Delta{\cal E}_D\,$
(if $\alpha<\beta$ the first term changes to $4\alpha k_0$).
Thus, the absorption bandwidth could be manipulated 
by tuning the coupling strength $\alpha$ and/or through
variations of the electron density $n=k_0^2/2\pi$.
This fact could also be used to determine the sign
of $\alpha-\beta$.\cite{cond-mat}
\begin{figure}
\includegraphics[width=0.5\textwidth]{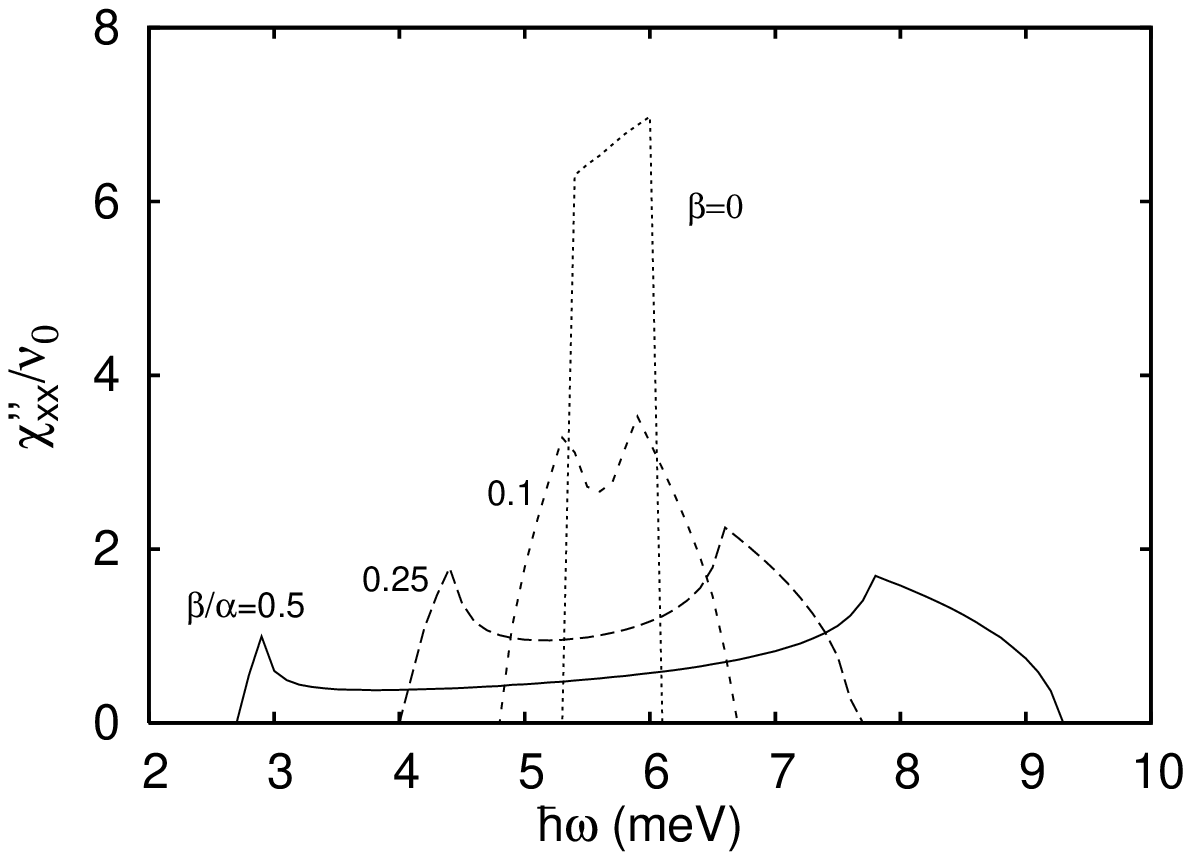}
\caption{Imaginary part of the spin susceptibility $\chi_{xx}(\omega)$
for several values of the ratio $\beta/\alpha$. Other parameters are
as those used in Fig.\,1\,.}
\end{figure}

To understand the structure of the spectra of Fig.\,2, we
further note that, according to eq. (\ref{Im_chi}),
the minimum (maximum) photon energy $\hbar\omega_+\,$($\hbar\omega_-$)
required to induce optical transitions between the initial $\lambda=-$ 
and the final $\lambda=+$ subband corresponds to the excitation of an 
electron with wave vector lying on the $q_+\,$($q_-$) Fermi line at 
$\theta_+=\pi/4$ or $5\pi/4$ ($\theta_-=3\pi/4$ or $7\pi/4$), 
giving $\hbar\omega_{\pm}=\hbar\Omega_{\pm}(\theta_{\pm})=
2k_0|\alpha \mp \beta|\mp 2m^*(\alpha \mp \beta)^2/\hbar^2$.
The absorption edges in the spectrum of Fig.\,2b correspond exactly to 
$\hbar\omega_{\pm}$. 
The function $\chi''_{\mu\mu'}(\omega)$ can also
be written as a line integral along the arcs of the 
resonant curve $C_r(\omega)$ lying within the region 
enclosed by the Fermi lines $q_{\lambda}(\theta)$; see Fig.\,1.
The peaks observed in Fig.2b correspond to electronic excitations involving 
states with allowed wave vectors on $C_r(\omega)$ such that 
$|\nabla_{\bf k}(\varepsilon_+-\varepsilon_-)|$ takes its minimum
value. The first (second) peak is at a photon energy $\hbar\omega_a$ 
($\hbar\omega_b$) for which the major (minor) semi-axis of the ellipse
$C_r(\omega)$ (Fig.\,1) coincides with the Fermi line $q_-(\theta_+)$
$(q_+(\theta_-))$, hence $\hbar\omega_a=\hbar\Omega_-(\theta_+)=2k_0|\alpha-\beta|+
2m^*(\alpha-\beta)^2/\hbar^2$ and $\hbar\omega_b=\hbar\Omega_+(\theta_-)
=2k_0|\alpha+\beta|-2m^*(\alpha+\beta)^2/\hbar^2$. The spectrum of
$\chi''_{xx}(\omega)$ looks very similar to the joint density of
states for the spin-split bands $\varepsilon_{\pm}$.\cite{cond-mat}
The unequal splitting at the 
Fermi level along the symmetry $(1,1)$ and $(-1,1)$ directions is thus 
responsible for the peaks at photon energies $\hbar\omega_a$ and 
$\hbar\omega_b$ respectively, giving meaning to the structure of the 
spectrum.
The overall magnitude 
and the asymmetric shape of the spectrum are due to the 
factor $g_{\mu\mu'}(\theta)/\Delta^4(\theta)$ in eq.(\ref{Imchi}). 
The results for several values of $\beta/\alpha$ are shown in Fig.\,3.

The real part of $\chi_{\mu\mu'}(\omega)$ presents additional spectral
features. For photon energies in the range $\hbar\omega_a\leq
\hbar\omega\leq\hbar\omega_b$ we find numerically that it
takes the constant values $\chi'_{xx}(\omega)=\nu_0/2$ and
 $\chi'_{xy}(\omega)=-(\nu_0/2)(\alpha^2+\beta^2)/2\alpha\beta$.
The spectral characteristics of the response displayed in Fig.\,2a 
shows that the magnitude and the
direction of the dynamic spin magnetization could be modified
via electrical gating and/or by adjusting the exciting frequency.
This suggests new possibilities of electrical manipulation of the 
spin orientation in a 2DEG in the presence of competing Rashba and 
Dresselhaus SO couplings.

Following Ref.\,\cite{Loss04} we have also obtained the static
value of $\chi_{\mu\mu'}(\omega)$ for finite momentum relaxation rate $\eta>0\,$
(see eq. (\ref{chi_ij})). This parameter accounts phenomenologically for
dissipation effects due to impurity scattering. We found that, 
to linear order in $\varepsilon_{R,D}/\hbar\eta$, it vanishes as
($\alpha\neq\beta\neq 0$)
\begin{equation}
\chi_{xx}(0;\eta)\approx 4\nu_0\,\left(\frac{\varepsilon_F}
{\hbar\eta}\right)\,\left(\frac{\varepsilon_R+\varepsilon_D}
{\hbar\eta}\right) \hspace*{1.3cm}
\chi_{xy}(0;\eta)\approx -4\nu_0\,\left(\frac{\varepsilon_F}
{\hbar\eta}\right)\,\frac{2\sqrt{\varepsilon_R\varepsilon_D}}
{\hbar\eta} \ \ .
\end{equation} 
It is also possible to relate the spin current response to
the spin density response. The definition of the
spin conductivity $\sigma^{s,z}_{iy}(\omega)$ describing a 
$z-$polarized-spin current flowing in the $i-$direction as a response
to the field $E(\omega){\bf\hat{y}}$ involves the commutator
$[{\cal J}^z_i(t),j_y(0)]$, where $j_i=ev_i$ and ${\cal J}^z_i=
\hbar\{\sigma_z,v_i\}/4$ are the charge and spin current operators,
respectively. Using the velocity operator
${\bf v}({\bf k})=\nabla_{\bf k}H/\hbar=\hbar{\bf k}/m^* + 
{\bf\hat{x}}(\beta\sigma_x+\alpha\sigma_y)/\hbar
-{\bf\hat{y}}(\alpha\sigma_x+\beta\sigma_y)/\hbar$, this commutator
can be written in terms of the correlators 
$[\sigma_i(t),\sigma_j(0)]$, ($\,i=x,y$), which determines the spin 
susceptibility (\ref{Kubo_chi}). Thus, the following relations can be
derived

\begin{equation}
\frac{\sigma^{s,z}_{xy}(\omega)}{e/8\pi}=\left(
\frac{\alpha^2+\beta^2}{\alpha^2-\beta^2}\right)\,
\frac{\chi_{xx}(\omega)}{\nu_0/2}+\left(
\frac{2\alpha\beta}{\alpha^2-\beta^2}\right)\,
\frac{\chi_{xy}(\omega)}{\nu_0/2}
\end{equation}
\begin{equation}
\frac{\sigma^{s,z}_{yy}(\omega)}{e/8\pi}=\left(
\frac{2\alpha\beta}{\alpha^2-\beta^2}\right)\,
\frac{\chi_{xx}(\omega)}{\nu_0/2}+\left(
\frac{\alpha^2+\beta^2}{\alpha^2-\beta^2}\right)\,
\frac{\chi_{xy}(\omega)}{\nu_0/2} \ \ .
\end{equation}

These expressions are formally equivalent to eqs. (39) and (40) 
of Ref.\,\cite{Erlingsson}.
This connection is very
convenient because a spin polarization is more experimentally
accesible than a spin current.

In summary, we have calculated the finite frequency  spin susceptibility 
tensor of a two-dimensional electron gas with competing Rashba and Dresselhaus 
spin-orbit interaction. We find that the angular anisotropy of the energy 
spin-splitting introduced by the interplay  between both SO coupling strengths 
yields a finite-frequency response with 
spectral features that are  significantly different from that of a pure 
Rashba (Dresselhaus) coupling  case. As a consequence, an optically modulable 
spin density response is then achievable in such systems which may be useful 
for spintronics applications. 

\newpage

This work was supported by CONACyT-Mexico 
grants J40521F, J41113F, and by DGAPA-UNAM IN114403-3.

\end{document}